\documentclass[10pt,conference,a4paper,twocolumn]{IEEEtran}

\IEEEoverridecommandlockouts 

\usepackage{cite}
\usepackage{amsmath,amssymb,amsfonts}
\usepackage{algorithmic}
\usepackage{graphicx}
\usepackage{textcomp}
\usepackage{hyperref}
\usepackage{booktabs}
\usepackage[ruled,linesnumbered]{algorithm2e}
\usepackage[dvipsnames]{xcolor}
\def\BibTeX{{\rm B\kern-.05em{\sc i\kern-.025em b}\kern-.08em
    T\kern-.1667em\lower.7ex\hbox{E}\kern-.125emX}}
   
\usepackage{xspace}
\makeatletter
\DeclareRobustCommand\onedot{\futurelet\@let@token\@onedot}
\def\@onedot{\ifx\@let@token.\else.\null\fi\xspace}

\def\eg{\emph{e.g}\onedot} 
\def\ie{\emph{i.e}\onedot} 
 
\def\etc{\emph{etc}\onedot}

\makeatother

\makeatletter
\def\ps@IEEEtitlepagestyle{%
  \def\@oddfoot{\mycopyrightnotice}%
}
\def\mycopyrightnotice{%
\begin{minipage}{\textwidth}
\centering \footnotesize
Copyright~\copyright~2023 IEEE. Personal use of this material is permitted.  Permission from IEEE must be obtained for all other uses, in any current or future media, including reprinting/republishing this material for advertising or promotional purposes, creating new collective works, for resale or redistribution to servers or lists, or reuse of any copyrighted component of this work in other works.
\end{minipage}
}
\makeatother

\newcommand{\mname}[0]{\textit{Split-Et-Impera}}
\newcommand{\isplit}{\textsc{I-Split}\xspace}
\newcommand{\cs}{\textsc{CS}\xspace}
\newcommand{\gradcam}{Grad-CAM\xspace}

\usepackage{soul}
\setulcolor{Red}

\begin{document}

\title{
Split-Et-Impera: A Framework for the Design of Distributed Deep Learning Applications
\thanks{
This project has received funding from the European Union’s Horizon 2020 research and innovation programme under the Marie Sklodowska-Curie grant agreement No. 894237, and has been also partially supported by the Italian Ministry of Education, University and Research (MIUR) with the grant ``Dipartimenti di Eccellenza'' 2018-2022, and by Fondazione Cariverona with the grant ``Ricerca \& Sviluppo''.
}
}

\author{
\IEEEauthorblockN{Luigi Capogrosso, Federico Cunico, Michele Lora, Marco Cristani, Franco Fummi, Davide Quaglia}
\IEEEauthorblockA{\textit{Department of Computer Science, University of Verona, Italy}}
{\tt name.surname@univr.it}
}

\maketitle
\IEEEpubidadjcol

\begin{abstract}
Many recent pattern recognition applications rely on complex distributed architectures in which sensing and computational nodes interact together through a communication network. Deep neural networks (DNNs) play an important role in this scenario, furnishing powerful decision mechanisms, at the price of a high computational effort. Consequently, powerful state-of-the-art DNNs are frequently split over various computational nodes, \eg{}, a first part stays on an embedded device and the rest on a server. Deciding where to split a DNN is a challenge in itself, making the design of deep learning applications even more complicated. Therefore, we propose \mname{}, a novel and practical framework that \textit{i)} determines the set of the best-split points of a neural network based on deep network interpretability principles without performing a tedious try-and-test approach, \textit{ii)} performs a communication-aware simulation for the rapid evaluation of different neural network rearrangements, and \textit{iii)} suggests the best match between the quality of service requirements of the application and the performance in terms of accuracy and latency time. 
\end{abstract}


\section{Introduction}
\label{sec:introduction}

\begin{figure}[t]
\begin{center}
\includegraphics[width=.95\linewidth]{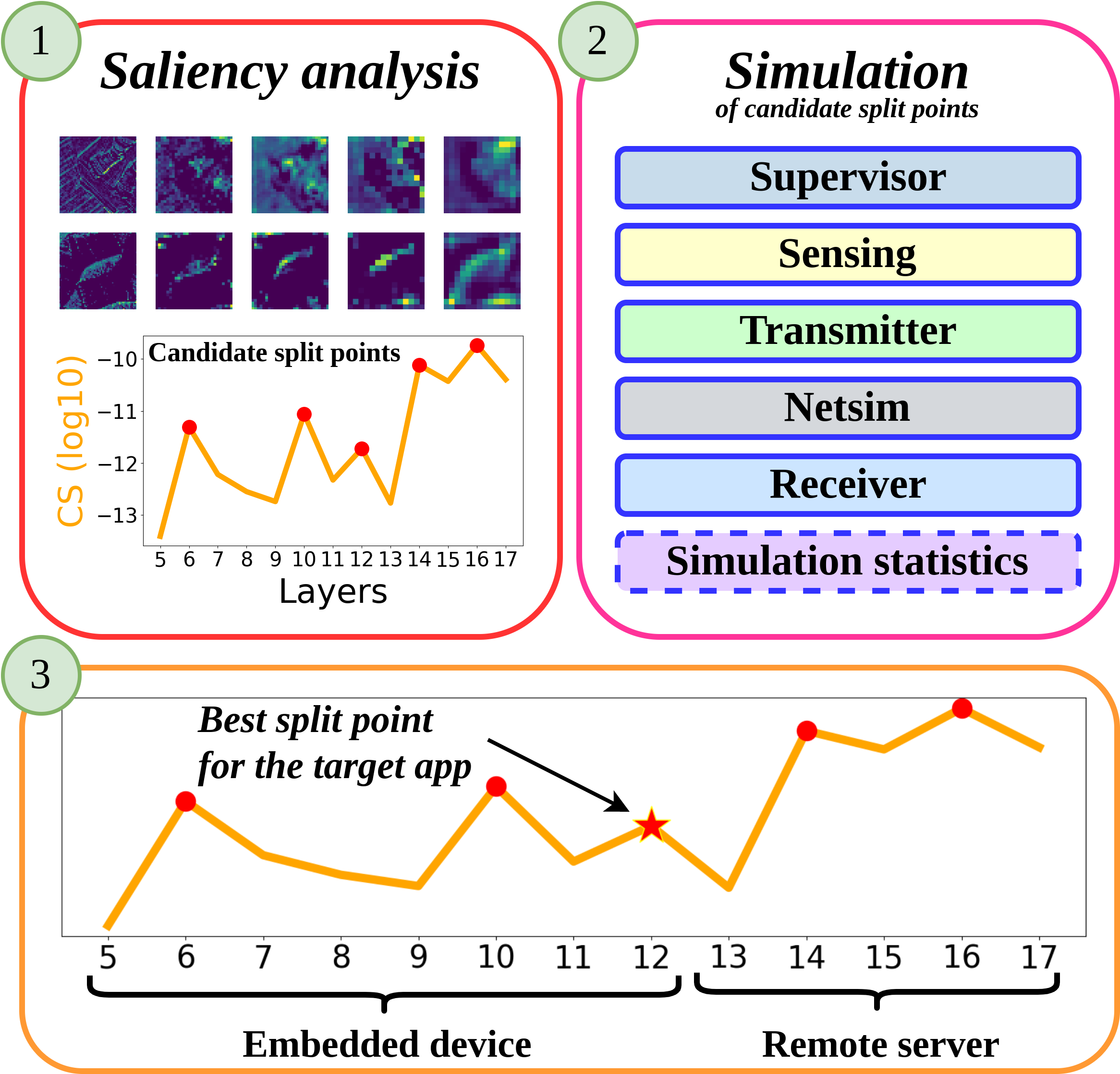}
\end{center}
\caption{\textit{The \mname{} framework}: \textit{i)} Determine the saliency-based split point candidates. The inputs are fed into a neural network to extract the saliency maps using the \gradcam{} algorithm at each layer. Then, we average over all the maps to generate the final \textit{Cumulative Saliency} (\cs{}) curve with the candidate split points. \textit{ii)} Simulate each candidate by reproducing its computation and its transmission. \textit{iii)} Use the simulation statistics to find the best-split point satisfying the application constraints on accuracy and latency.}
\label{fig:Schema}
\end{figure}

In the last decade, DNNs achieved state-of-the-art performance in a broad range of problems~\cite{abiodun2018state}, \eg{}, object classification~\cite{zhao2017survey} and feature detection~\cite{li2015survey}, spanning from smart building~\cite{capogrosso2022toward} to healthcare~\cite{skenderi2021dohmo}. The flexibility that makes DNNs such a pervasive technology comes at a price: the computational requirements of some DNNs preclude their deployment on most of the resource-constrained devices available today. A possible alternative is running simplified models, such as MobileNetV2~\cite{sandler2018mobilenetv2}, on the devices; we refer to this scenario as \textit{local-only computing} (LC)~\cite{matsubara2021split}. 
However, using simplified models negatively affects the overall accuracy.
Therefore, the solution that is most widely used to date is usually referred to as \textit{remote-only computing} (RC)~\cite{shi2012serendipity}, which consists in transferring the data captured by the device to a high-performance machine through a communication network and then sending back the results to the device if needed. In this case, the communication network may become a bottleneck and it should be properly configured to match the quality of service requirements of the application. \textit{Split computing} (SC) has been recently proposed~\cite{kang2017neurosurgeon} to divide the DNN model into a ``head'' on the sensing device and a ``tail'' on the remote server: the output of the last layer of the head is transmitted over the communication network as input for the tail.

Therefore, the design of distributed deep learning applications consists in exploring a three-dimensional design space. Indeed, the final implementation will be determined by the choices of the \textit{computation platform}, the \textit{communication architecture}, and the \textit{DNN}.
Whereas the first dimension is deterministic, in the sense that the choice of a specific platform produces certain performances,
dealing with communication networks and DNNs introduces uncertainty. A DNN is a statistical classifier with millions of parameters whose timing is non-deterministic, as it is its efficacy that is usually measured in terms of accuracy. It follows that a given DNN can be evaluated only by training and testing it on some validation partition. Regarding SC scenarios, the situation becomes even harder since the splitting technique typically requires specific training~\cite{matsubara2022bottlefit}, while manipulating diverse SC configurations requires days~\cite{matsubara2021split}.


This work shows that the optimal split point to match the quality of service (QoS) requirements of an application depends on the \textit{saliency} of the network layers. 
This idea lowers the design and deployment time from days to hours, offering a 
practical developing framework grounded on a solid theory. 
Each layer interrogates its input about specific signal properties, such as the shape of the objects contained in an image. Some of the properties are crucial (\ie{}, \textit{salient}) for the final classification. For example, if a DNN has to recognize different shapes, the information flows have to be preserved until the shape layer; after that, a split can be carried out. 

As a major result, this work shows that split locations must account not only for \textit{dense}~\cite{sbai2021cut} but also for \textit{informative} data. The rationale is to use a saliency-based splitting point search~\cite{cunico2022split} to preserve the portions of the network where crucial decisions are taken. We further generalize the theory in~\cite{cunico2022split}. Then, on top of the theory, we propose \mname{}\footnote{\url{https://github.com/luigicapogrosso/split_et_impera}} (Figure~\ref{fig:Schema}): a fast and user-friendly framework that eases the design of a distributed architecture executing one or more DNNs. Other than accurately mimicking diverse communication protocols and application requirements, \mname{} introduces a unique feature: it suggests the proper configuration to match the application's QoS requirements and provide optimal performance in terms of accuracy and latency time. Furthermore, since manipulating diverse SC configurations may require days of computation, \mname{} allows eliminating several configurations through communication-aware simulations. 

\mname{} has been applied to simulate a classification task within a real Industry 4.0 scenario, using the LC, RC, and SC settings. It allows finding a set of possible split points for DNNs and deciding which is the best design to match the QoS requirements through a simulation. Furthermore, we used \mname{} to evaluate the application design and transmission protocol selection by modeling the application's transmission details. Experiments show the effectiveness of the proposed frameworks.


\section{Related Work}
\label{sec:related}

This section provides a concise overview of the main architectural typologies used for distributed deep-learning applications. Then, we introduce the state-of-the-art on simulation for distributed DNN architectures.


\subsection{Distributed deep learning architectures}
We focus on architectures operating through a DNN model $M(\cdot{})$, whose task is to produce the inference output $y$ from an input $x$. 
Three types of architectures used for distributed deep learning applications can be identified in the literature, \ie{}, \textit{LC}, \textit{RC}, and \textit{SC}.

\paragraph*{Local computing}
under this policy, the entire computation is performed on the sensing devices, \eg{}, mobile phones, cameras equipped with micro-controller, \etc{} Therefore, the function $M(x)$ is entirely executed by the edge device.

This approach is well-suited for applications characterized by a high data transfer rate, such as automated video analysis~\cite{laureshyn2010application}. It provides low latency since the computing element is very close to the sensor. On the other hand, it does not fit with DNN-based architectures requiring powerful hardware. Usually, simpler DNN models $\bar{M}(x)$ that use specific architectures (\eg{}, depth-wise separable convolutions) are used to build lightweight networks, such as MobileNet~\cite{sandler2018mobilenetv2}. 


\paragraph*{Remote computing}
the input $x$ is transferred through the communication network and then is processed at the remote system through the function $M(x)$.

This architecture preserves full accuracy considering the higher power budget of the remote system, but leads to high latency and consuming bandwidth due to the input transfer~\cite{kim2009cloud}.

\paragraph*{Split computing}
The SC paradigm divides the DNN model into a head, executed by the sensing device, and a tail executed by the remote system. It combines the advantages of both LC and RC thanks to the lower latency and, more importantly, drastically reduces the required transmission bandwidth by compressing the input to be sent $x$ through the use of an autoencoder~\cite{matsubara2021split}. We define the encoder and decoder models as $z_{l}=F(x)$ and $\bar{x}=G(z_{l})$, which are executed at the edge, and remotely, respectively. The distance $d(x,\bar{x})$ defines the performance of the encoding-decoding process.


In this paper, we consider the theory of~\cite{cunico2022split} (\isplit{}) as a way to suggest which split configuration has to be preferred while avoiding an exhaustive model search. Specifically, our method is closely related to \isplit{} but has three major differences: \textit{i)} we have different motivations, \ie{}, our work is motivated by the proposal of a novel framework that can take into account aspects such as computation time, transmission topology, and protocol, that in \isplit{} are not presented. \textit{ii)} The \isplit{} approach is limited to working on images. Meanwhile, our method generalizes it to be used with any type of signal. \textit{iii)} Our framework performs a communication-aware simulation greatly reducing the time required for the evaluation of SC applications.

\subsection{Simulators for distributed DNN architectures}
A mobile computing system, based on several advanced partition schemes (\eg{}, BODP, MSCC), to enable parallel computation of DNN on mobile platforms is presented in~\cite{mao2017modnn}.

In~\cite{zhao2018deepthings}, is presented a framework for adaptively distributed execution of CNN-based inference. The focus is on the early layers, which largely contribute to the overall latency.

In~\cite{huang2019deepar}, the authors proposed a framework focused on a layer-level partitioning strategy to further improve the overall inference performance in an edge computing scenario.

In~\cite{alwasel2021iotsim}, the authors presented a simulation framework for analyzing and validating osmotic computing. Osmotic computing provides a model for the deployment of IoT in the integrated edge-cloud environment.

None of these simulators is considering all four elements which give a reliable blueprint of the final system (computation time, transmission topology, protocol, and accuracy). Therefore, to the best of our knowledge, we are the first in this respect. Most importantly, we are the first in providing suggested configurations. When using \mname{}, the engineer will just need to select the type of network to be used and its initial parameters (\ie{}, the model has to be trained). Then, the simulator can focus on a few selected alternatives to evaluate, avoiding exhaustive model state searches.

\section{Methodology}
\label{sec:saliency}

Splitting a DNN into multiple possible locations is the first step for identifying alternative distributed architectures to be simulated. The positions will be ranked according to the accuracy that the system will presumably achieve with each split. Then, the ranking will be simulated to provide the configuration design meeting the QoS requirements. 

The saliency-based splitting point search is implemented by using \gradcam{}~\cite{selvaraju2017grad}, which has been proved to pass the sanity check for saliency-based interpretability approaches~\cite{adebayo2018sanity}. Thus, ensuring that the saliency maps of \gradcam{} depend on the specific model instance taken into account.
Specifically, the saliency computation can be applied at every layer of size $n\times m \times z$ of a DNN model. The result is a $n \times m$ class activation map. Then, the mean of the class activation map is performed, giving a single value per layer. The values obtained by all the layers compose the \textit{Cumulative Saliency} (\cs{}) curve for the trained model.

We assume that the DNN model is pre-trained. The \cs{} curve is computed on some testing set, composed of $C$ classes, each formed by $N_C$ input observations. For each $c$-th class, we proceed as follows.

For a given $i$-th layer of the neural network ($i=1,\ldots,I$), for each $j$-th input data belonging to class $c \in C$, we extract the class-discriminative activation map $L^{i}_{j,c}$. We start by computing the feature map importance coefficient $\alpha{}_{i,j}^{c}$:
\begin{equation}
\alpha{}_{i,j}^{c}=\frac{1}{z}\sum_{n}^{}\sum_{m}^{}\frac{\partial y^{c}}{\partial F_{n,m}^{i,j}}
\end{equation}
where $F^{i,j}\in R^{n \times m\times z}$ is the feature map of the convolutional layer $i$ for the input $j$, and $y^{c}$ is the output for the class $c$.

The weight $\alpha{}_{i,j}^{c}$ represents a partial linearization of the deep network downstream from $F$ and captures the value of the feature map of the $i$-th layer for a target class $c$. At this point, we perform a weighted sum between the value just calculated and the feature maps $F^{i,j}$ of the chosen layer. Finally, the \textit{ReLU} activation function is applied to reset the negative values of the gradient to zero obtaining the class activation map $L^{i}_{j,c}$ for a specific query input $j$:
\begin{equation}
L^{i}_{j,c}=ReLU\left(\sum_{k=i}^{I}\alpha{}_{k,j}^{c}F^{k,j}\right)
\end{equation}
This map focuses on the $i$-th layer, summing from the class activations of the networks back until $i$. 
We aim to obtain a single value for our class activation map $L^{i}_{j,c}$. Thus, we simply sum over the dimensions of $F^{i,j}\in R^{n \times m\times z}$, obtaining the per-input saliency values $\cs{}_{j,c}^{i}$. Ideally, computing these values for each $i$-th layer of the network gives a curve showing how the input has triggered the different layers of the network. At this point, averaging over all the inputs of all the classes provides the final CS curve, where the $i$-th point $\cs{}^i$ of the curve is a surrogate of the information conveyed through the $i$-th layer towards the decision for the correct class, for all the classes into play.

Once the \cs{} curve has been computed, the candidate split points can be identified by the layers that give local \cs{} maxima. Given these $N$ points, we will then go on to simulate each of them, and we choose the one with the best performance concerning the QoS requirements of the application.

Let $T^{i}$ be the target layer for splitting the model at index $i$ and $T^{i+1}$ the subsequent layer. We divide the network into three main blocks: \textit{i)} the head, running on the edge device, is composed of the first layers of the original DNN architecture, up to layer $T^i$; \textit{ii)} the bottleneck, an undercomplete autoencoder that learns low-dimensional latent attributes which explain the input data; \textit{iii)} and the tail, from layer $T^{i+1}$ to the very end of the network, that is executed on the server-side. The encoder part of the bottleneck is deployed to the edge device, while the decoder is executed on the server side.

In order to train the entire model $M(\cdot{})$, we first train our bottleneck. Then we fine-tune the entire network end-to-end. We create an undercomplete autoencoder which acts as a bottleneck inserted after $T^{i}$. The use of bottlenecks is standard in the SC literature~\cite{matsubara2021split}. This bottleneck should learn to replicate the input, which is the feature map in output at layer $T^{i}$. Therefore, given $\{\boldsymbol{I}_j, j=1,2,...,n\}$ as the $n$ number of training data, we train the sole bottleneck freezing the remaining network with the following loss function:
\begin{equation}
\mathcal{L}_{AE} = \frac{1}{n} \sum_{j=1}^{n} || \Phi_{T^{i}}(\boldsymbol{I}_j) - \Psi(\Phi_{T^{i}}(\boldsymbol{I}_j); \boldsymbol{W}_{AE})||^2
\end{equation}
with $\Phi_{T^{i}}$ the model layers up to the $i$-th layer $T^{i}$ (target layer), and $\Psi$ the AE part of the model with weights $\boldsymbol{W}_{AE}$.

After training the bottleneck, we perform a fine-tuning of the network with the following loss function:
\begin{equation}
\mathcal{L}_{task} = \frac{1}{n} \sum_{j=1}^{n} || \Phi_M(\boldsymbol{I}_j;\boldsymbol{W}_M) - \hat{y}_j ||^2
\end{equation}
with $\Phi_M$ the full $M$'s DNN layers, $\boldsymbol{W}_{M}$ the weights of the model $M$, and $\hat{y}_j$ the ground truth label for the input $j$.

\section{simulator architecture}
\label{sec:simulator}

The architecture of \mname{} is presented in Figure~\ref{fig:Schema}. To offer a large model state space, it is divided into five main layers: \textit{supervisor}, \textit{sensing}, \textit{transmitter}, \textit{netsim}, and \textit{receiver}. The network simulation is based on the SCNSL~\cite{fummi2008systemc} library, while the simulator is implemented in Python.


The review provided in Section~\ref{sec:related} led us to aim at developing a communication-aware simulation platform for distributed deep learning applications that accurately emulates the behavior and the timing of the computation, communication, and inference performed by the system. The simulator has to be \textit{modular}, \textit{portable}, and \textit{language independent}, \ie{}, the functional blocks of the framework should be integrated into the model independently of their implementation programming language. Furthermore, it must allow the integration of real-world components, such as a real computing system: the so-called ``hardware-in-the-loop'' (HIL)~\cite{bullock2004hardware}.


The \textit{supervisor} controls all the events and operations happening during the simulations. The \textit{sensing} layer is a high-level wrapper encoding the application into the architecture. The \textit{transmitter} module is responsible for the XMTR. The \textit{netsim} must guarantee that the simulation emulates the behavior of an actual communication channel: it has to execute events in the correct temporal order while taking into account the physical features of the channel, such as propagation delay, and signal loss. Finally, the \textit{receiver} is responsible for the RCVR.

Specifically, \mname{} requires the following inputs: 
\begin{enumerate}
\item \textit{Test scenario}: LC, RC, or SC (as described in Section~\ref{sec:related}).
\item \textit{Training set}: the set of data used to train the model and make it learn the hidden features/patterns in the data.
\item \textit{Trained instance of a DNN}: the network parameters, once it has been trained on a general purpose processor.
\item \textit{Test set}: the data with which the simulation is run, a proxy of the real setting the framework will work on.
\item \textit{Communication network modeling}: through some parameters which will be explained in the following.
\end{enumerate}
The communication network modeling parameters are:
\begin{enumerate}
\item \textit{Communication protocol}: the transport layer protocol that must be used, \ie{}, TCP or UDP.
\item \textit{Channel latency}: the time that each packet spends to travel from the sender to the receiver, \eg{}, $100~\mu{}s$.
\item \textit{Channel capacity}: the link available bandwidth. 
\item \textit{Interface speed}: to better model different hardware devices, even the physical speed can be set to match the desired target hardware, \eg{}, $1000~Mb/s$ to represent a Gigabit connection, $100~Mb/s$ for Fast-Ethernet, $160~Mb/s$ to represent Wi-Fi, \etc{}
\item \textit{Saboteur}: the network loss rate.
\end{enumerate}
%

The output of \mname{} is of two types: \textit{i)} the suggested configurations to simulate, ranked by the classification accuracy that the network is assumed to achieve. The engineer may then decide to simulate all or only a subset of them. \textit{ii)} The simulation results of the configurations selected in the previous step to deploy the application using the best design.

\section{Experimental Results}
\label{sec:experimental}

In this section, we show how the design of a typical real application can be carried out by \mname{}. In particular, we focus on the classification task of children's toys (such as boats, airplanes, \etc{}) passing on a conveyor belt within a real Industry 4.0 scenario: the ICE Laboratory located in Verona\footnote{\url{https://www.icelab.di.univr.it/}.}. 

As DNN, we use the PyTorch implementation of the VGG16~\cite{simonyan2014very}. Other than being the workhorse of the industrial computer vision and pattern recognition~\cite{liu2021toward}, VGGs has recently been rejuvenated with diverse ``cousin'' extensions, like the RepVGG~\cite{ding2021repvgg}, a VGG-like inference-time body composed of nothing but a stack of $3\times{}3$ convolution and ReLU. RepVGG models run faster than ResNet-50 or ResNet-101~\cite{he2016deep} with higher accuracy and show a favorable accuracy-speed trade-off compared to the state-of-the-art models like EfficientNet~\cite{tan2019efficientnet}. Other than these reasons, VGG suits very well for SC: in its classic form, it exhibits a good base accuracy, a large number of parameters, is relatively deep in terms of the number of layers, and the trend of the layers in terms of complexity and output size is not regular, making the split point determination not trivial.

For the image classification task, we train our model on the CIFAR10 dataset~\cite{krizhevsky2009learning} up to 20 epochs with a learning rate of $5\times{}10^{-3}$, using Adam~\cite{kingma2014adam} as optimizer. We tested our application on images acquired in the ICE Lab. CIFAR10 has to be considered as a placeholder for bigger datasets (\ie{}, ImageNet~\cite{deng2009imagenet}); nonetheless, the focus here is to show the potentialities of the \mname{} and not beating the state-of-the-art in a specific computer vision challenge. 

In all the experiments, the split has been realized by placing an undercomplete autoencoder with a 50\% of compression rate. For the training of the encoder/decoder pair, we run up 50 epochs with a learning rate of $5\times{}10^{-4}$, always using Adam as optimizer.

\subsection{Saliency-based split point search}
\begin{figure}[t]
\begin{center}
\includegraphics[width=.95\linewidth]{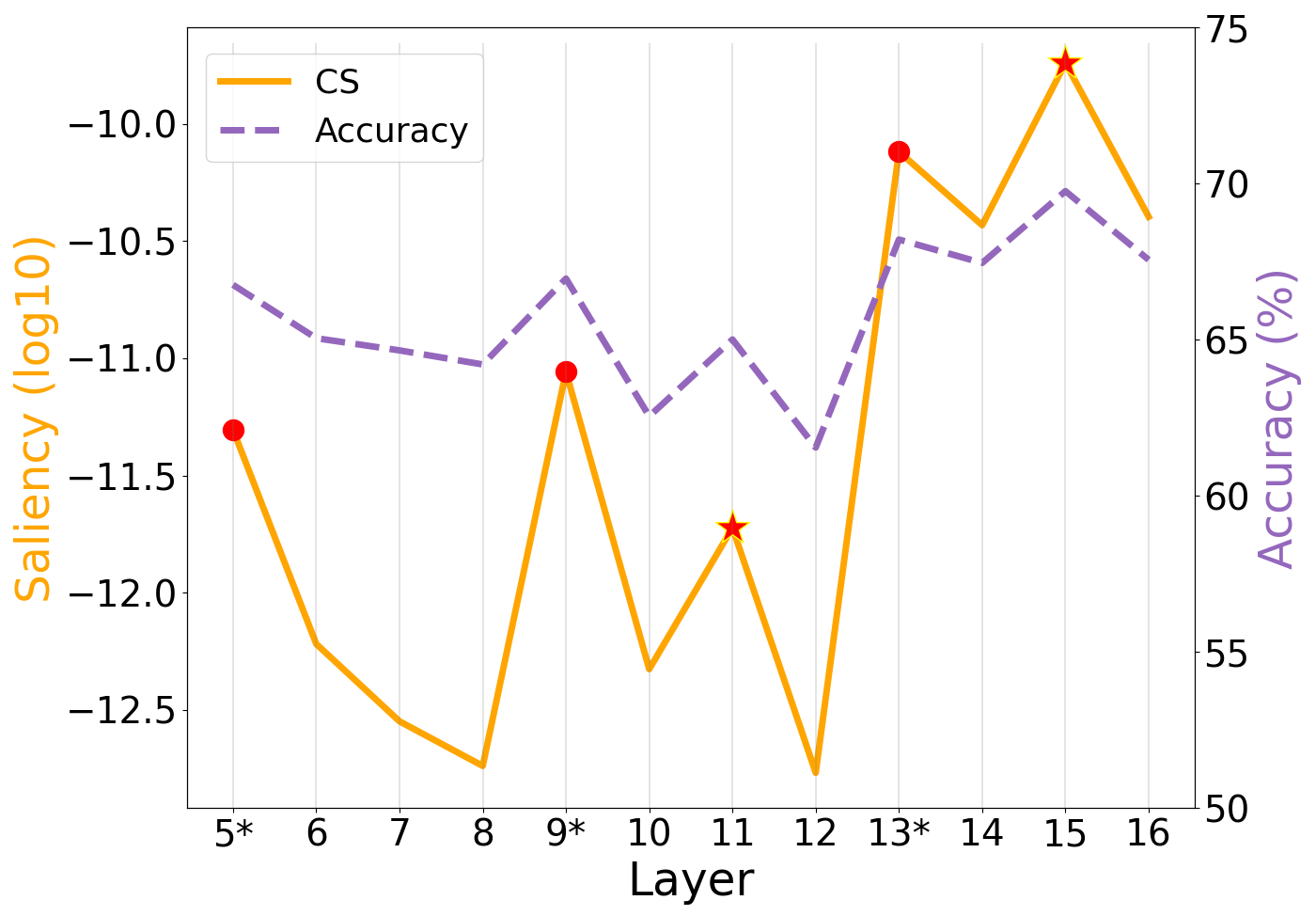}
\end{center}
\caption{\textit{Cumulative Saliency} (\cs{}) as a function of the layer compared with the accuracy of the DNN split in that layer. The peaks of the \cs{} curve correspond to the points in which accuracy is preserved despite split injection. Thus, the layers in which \cs{} has a local maximum are the best candidates for splitting. (*) represents a max pooling layer.} 
\label{fig:SEI_and_accuracy_VGG16}
\vspace{-1em}
\end{figure}

The first output of \mname{} is a set of candidate split points that are worth being simulated. Figure~\ref{fig:SEI_and_accuracy_VGG16} shows the results of the saliency-based split point search. 

We can clearly appreciate how the \cs{} saliency approach identifies as candidate split points (in red) layers $5$, $9$ and $13$, corresponding to \textit{block2\_pool}, \textit{block3\_pool} and \textit{block4\_pool} (dense data), and two additional points at layers $11$ and $15$ (informative data), corresponding to \textit{block4\_conv2} and \textit{block5\_conv2}, respectively. It is worth noting how layers with the same dimensionality, \ie{}, convolutional layers belonging to the same block, do not express the same importance as shown by the \cs{} curve. In these regions of uniform dimensionality, we select layers that have higher importance, and we will show that the \cs{} values directly translate into higher classification accuracy. Figure~\ref{fig:SEI_and_accuracy_VGG16} confirms that \cs{} is a good proxy for the overall classification accuracy and thus it is worth splitting the network into the layers in which \cs{} exhibits a local maximum.

Given the results of the \cs{} computing procedure, which we remind is done without having retrained the network, then we can explore the complete behavior of the application, including its transmission aspects, just splitting into the candidate layers. In particular, due to the lack of space, in the next section, we highlight simulation results only after splitting at layers $11$ and $15$, as well as for RC.

\subsection{Communication-aware split point selection}
\begin{figure}[t]
\centering
\includegraphics[width=0.8\columnwidth]{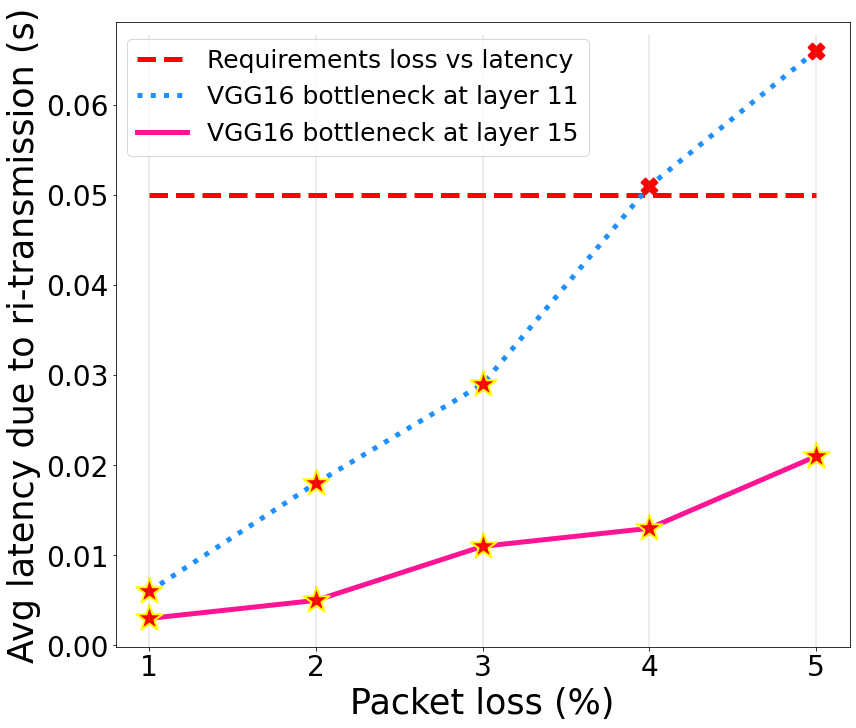}
\caption{Evaluation of the impact of the network on split point selection for the classification task of the images captured inside the ICE Lab. The network channel is 1 GB/s Full-Duplex with TCP protocol.}
\label{fig:SC2}
\end{figure}

Figure~\ref{fig:SC2} refers to the experiment of the SC scenario for the classification task of the images captured inside the ICE Lab. The network channel is 1 GB/s Full-Duplex with TCP protocol. The application presents a constraint on the maximum frame latency of 0.05~s (\ie{}, 20~FPS), given by the velocity of the conveyor belt. The split point is at layer $11$ and $15$, respectively. 

Figure~\ref{fig:SC2} highlights how the latency increases with the packet loss rate, due to TCP retransmission in case of packet loss. On the other hand, however, this preserves the accuracy of the application. In particular, the solid pink curve shows that with the split at layer $15$ the application requirements are always satisfied independently of the packet loss rate. Vice versa, the dotted blue curve shows that with the split at layers $11$, the 20~FPS constraint cannot be satisfied when the packet loss rate is more than 3\%. The behavior of the simulator meets expectation, \ie{}, by splitting the network at layer $11$, the amount of transmitted data is greater than the one obtained by splitting the network at layer $15$, and because of the retransmissions, the latency increases, up to violate the application constraints represented by the dashed red line.

This experiment shows the main claim of our work: given a set of possible split point solutions (see Figure~\ref{fig:SEI_and_accuracy_VGG16}), our framework through a rapid evaluation allows deciding which DNN configuration is compatible with the application requirements while considering the communication setup.

\subsection{Network protocol selection}

\begin{figure}[t]
\centering
\begin{minipage}[t]{0.49\linewidth}
\centering
\includegraphics[width=\linewidth]{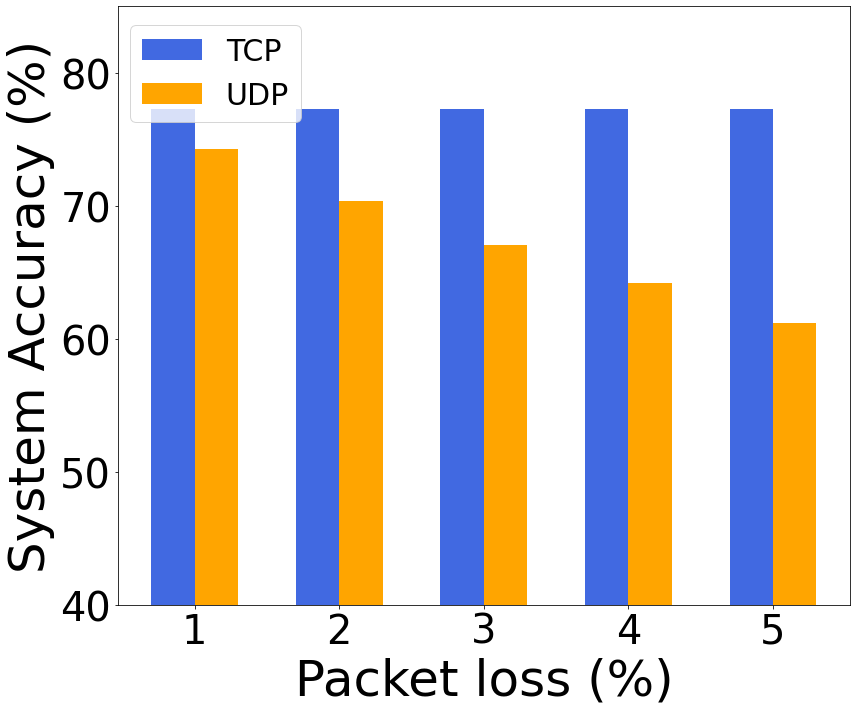}
\label{fig:CDE_and_SEI_VGG16_notMNIST}
\end{minipage}
\begin{minipage}[t]{0.49\linewidth}
\centering
\includegraphics[width=\linewidth]{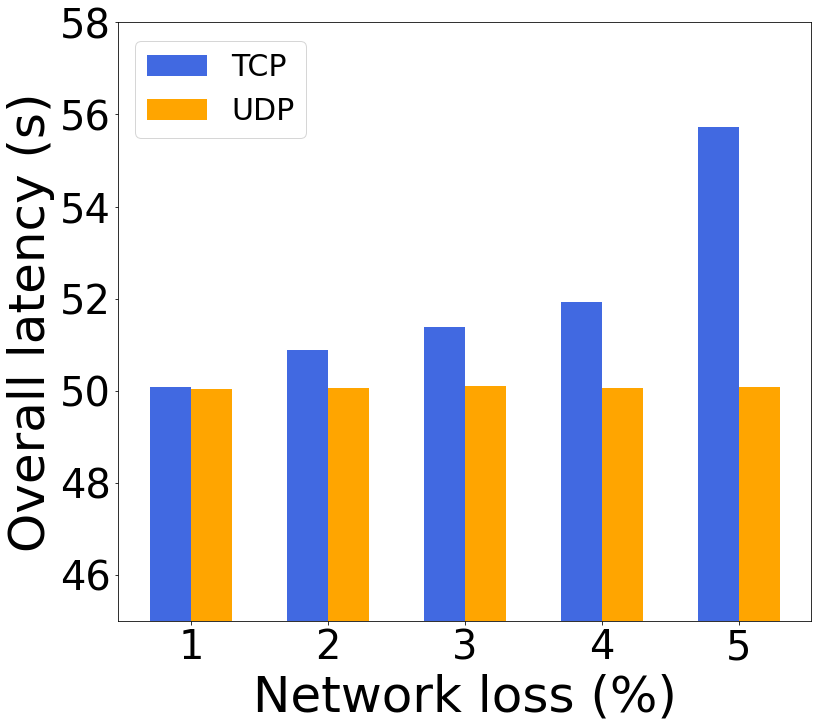}
\end{minipage}
\caption{Accuracy (left) and latency (right) results on the CIFAR10 test set for the classification task, contextualized in the RC scenario with the use of the TCP and UDP protocols. The network channel is 1 GB/s Full-Duplex.}
\label{fig:RC12}
\end{figure}

Figure~\ref{fig:RC12}-left and Figure~\ref{fig:RC12}-right show the system accuracy and the overall latency, related to the classification task in the full RC scenario, this time on the CIFAR10 test set, with the use of the TCP and UDP protocols. The network channel is 1 GB/s Full-Duplex.

Figure~\ref{fig:RC12}-left shows that, using TCP, the application accuracy does not depend on the packet loss rate. However, Figure~\ref{fig:RC12}-right shows that this feature comes at a cost: with TCP the overall latency is much greater, so it is required to make sure that this is compatible with the application requirements. UDP protocol shows a dual behavior, the latency is minimized and kept independent of the packet loss rate but the accuracy decreases in case of loss since no error checking and recovery services are provided. 

These experiments show that the proposed framework allows modeling an application's transmission details to jointly perform split point selection and transmission protocol selection.

\subsection{Neural network statistics}
\label{subsec:nnstats}
\begin{table}[t]
\centering
\caption{The neural network summary provided for the VGG16.}
\label{tab:nn_summary}
\begin{tabular}{lll}
\toprule 
\textbf{Layer} (type:depth-idx) & \textbf{Output Shape} & \textbf{Param} (\#) \\
\midrule
Sequential: 1-1 & [16, 512, 7, 7]       &   --          \\
Conv2d: 2-1     & [16, 64, 224, 224]    &   1.792       \\
ReLU: 2-2       & [16, 64, 224, 224]    &   --          \\
Conv2d: 2-3     & [16, 64, 224, 224]    &   36.928      \\
\dots{}         & \dots{}               &   \dots{}     \\
AdaptiveAvgPool2d: 1-2  & [16, 512, 7, 7]   &   --          \\
Sequential: 1-3         & [16, 1000]        &   --          \\
Linear: 2-32            & [16, 4096]        &   102.764.544 \\
ReLU: 2-33              & [16, 4096]        &   --          \\
\dots{}         & \dots{}               &   \dots{}     \\
Linear: 2-38            & [16, 1000]        &   4.097.000   \\
\bottomrule
\end{tabular}
\end{table}

\begin{table}[t]
\centering
\caption{The neural network statistics provided for the VGG16.}
\label{tab:nn_stats}
\begin{tabular}{ll}
\toprule 
\textbf{Statistic} & \textbf{Value} \\
\midrule
Total params                    &   138.357.544 \\
Trainable params                &   138.357.544 \\
Total mult-adds (G)             &   247.74      \\
Forward/backward pass size (MB) &   1735.26     \\
Estimated Total Size (MB)       &   2298.32     \\
\bottomrule
\end{tabular}
\end{table}



\mname{} can also provide several statistics about the DNN. In particular, Table~\ref{tab:nn_summary} shows the summary of each constituent DNN module. These values are particularly useful to check whether the layers are producing data correctly. Furthermore, the values may be used to guess where overfitting and computational bottlenecks are likely to occur. Table~\ref{tab:nn_stats} shows the DNN statistics. The more trainable parameters a model has, the more computing power will be needed. Specifically, these reports are particularly useful when working with a customized DNN for which statistical information is not available in the existing literature.

\section{Conclusions}
\label{sec:conclusions}

In this work, we have presented \mname{}, a novel framework for the design of distributed deep learning applications based on split computing. As a result, \mname{} suggests the proper configuration to match the quality of service requirements of the application and provide high performance in terms of accuracy and latency time. The experiments give an overview of specific features of \mname{} and show that the results obtained are very good, while 
extensive benchmarks will be the subject of future work.



\bibliographystyle{IEEEtran}
\bibliography{bibi}

\begin{thebibliography}{10}
\providecommand{\url}[1]{#1}
\csname url@samestyle\endcsname
\providecommand{\newblock}{\relax}
\providecommand{\bibinfo}[2]{#2}
\providecommand{\BIBentrySTDinterwordspacing}{\spaceskip=0pt\relax}
\providecommand{\BIBentryALTinterwordstretchfactor}{4}
\providecommand{\BIBentryALTinterwordspacing}{\spaceskip=\fontdimen2\font plus
\BIBentryALTinterwordstretchfactor\fontdimen3\font minus
  \fontdimen4\font\relax}
\providecommand{\BIBforeignlanguage}[2]{{%
\expandafter\ifx\csname l@#1\endcsname\relax
\typeout{** WARNING: IEEEtran.bst: No hyphenation pattern has been}%
\typeout{** loaded for the language `#1'. Using the pattern for}%
\typeout{** the default language instead.}%
\else
\language=\csname l@#1\endcsname
\fi
#2}}
\providecommand{\BIBdecl}{\relax}
\BIBdecl

\bibitem{abiodun2018state}
O.~I. Abiodun, A.~Jantan, A.~E. Omolara, K.~V. Dada, N.~A. Mohamed, and
  H.~Arshad, ``{State-of-the-art in artificial neural network applications: A
  survey},'' \emph{Heliyon}, vol.~4, no.~11, p. e00938, 2018.

\bibitem{zhao2017survey}
B.~Zhao, J.~Feng, X.~Wu, and S.~Yan, ``{A survey on deep learning-based
  fine-grained object classification and semantic segmentation},''
  \emph{International Journal of Automation and Computing}, vol.~14, no.~2, pp.
  119--135, 2017.

\bibitem{li2015survey}
Y.~Li, S.~Wang, Q.~Tian, and X.~Ding, ``{A survey of recent advances in visual
  feature detection},'' \emph{Neurocomputing}, vol. 149, pp. 736--751, 2015.

\bibitem{capogrosso2022toward}
L.~Capogrosso, G.~Skenderi, F.~Girella, F.~Fummi, and M.~Cristani, ``{Toward
  Smart Doors: A Position Paper},'' \emph{arXiv preprint arXiv:2209.11770},
  2022.

\bibitem{skenderi2021dohmo}
G.~Skenderi, A.~Bozzini, L.~Capogrosso, E.~C. Agrillo, G.~Perbellini, F.~Fummi,
  and M.~Cristani, ``{DOHMO: Embedded Computer Vision in Co-Housing
  Scenarios},'' in \emph{2021 Forum on specification \& Design Languages
  (FDL)}.\hskip 1em plus 0.5em minus 0.4em\relax IEEE, 2021, pp. 01--08.

\bibitem{sandler2018mobilenetv2}
M.~Sandler, A.~Howard, M.~Zhu, A.~Zhmoginov, and L.-C. Chen, ``{Mobilenetv2:
  Inverted residuals and linear bottlenecks},'' in \emph{Proceedings of the
  IEEE conference on computer vision and pattern recognition}, 2018, pp.
  4510--4520.

\bibitem{matsubara2021split}
Y.~Matsubara, M.~Levorato, and F.~Restuccia, ``{Split computing and early
  exiting for deep learning applications: Survey and research challenges},''
  \emph{ACM Computing Surveys (CSUR)}, 2021.

\bibitem{shi2012serendipity}
C.~Shi, V.~Lakafosis, M.~H. Ammar, and E.~W. Zegura, ``{Serendipity: Enabling
  remote computing among intermittently connected mobile devices},'' in
  \emph{Proceedings of the thirteenth ACM international symposium on Mobile Ad
  Hoc Networking and Computing}, 2012, pp. 145--154.

\bibitem{kang2017neurosurgeon}
Y.~Kang, J.~Hauswald, C.~Gao, A.~Rovinski, T.~Mudge, J.~Mars, and L.~Tang,
  ``{Neurosurgeon: Collaborative intelligence between the cloud and mobile
  edge},'' \emph{ACM SIGARCH Computer Architecture News}, vol.~45, no.~1, pp.
  615--629, 2017.

\bibitem{matsubara2022bottlefit}
Y.~Matsubara, D.~Callegaro, S.~Singh, M.~Levorato, and F.~Restuccia,
  ``{BottleFit: Learning Compressed Representations in Deep Neural Networks for
  Effective and Efficient Split Computing},'' \emph{arXiv preprint
  arXiv:2201.02693}, 2022.

\bibitem{sbai2021cut}
M.~Sbai, M.~R.~U. Saputra, N.~Trigoni, and A.~Markham, ``{Cut, Distil and
  Encode (CDE): Split Cloud-Edge Deep Inference},'' in \emph{2021 18th Annual
  IEEE International Conference on Sensing, Communication, and Networking
  (SECON)}.\hskip 1em plus 0.5em minus 0.4em\relax IEEE, 2021, pp. 1--9.

\bibitem{cunico2022split}
F.~Cunico, L.~Capogrosso, F.~Setti, D.~Carra, F.~Fummi, and M.~Cristani,
  ``{I-SPLIT: Deep Network Interpretability for Split Computing},'' in
  \emph{2022 26th International Conference on Pattern Recognition
  (ICPR)}.\hskip 1em plus 0.5em minus 0.4em\relax IEEE, 2022, pp. 2575--2581.

\bibitem{laureshyn2010application}
A.~Laureshyn, \emph{{Application of automated video analysis to road user
  behaviour}}.\hskip 1em plus 0.5em minus 0.4em\relax Lund University, 2010.

\bibitem{kim2009cloud}
W.~Kim, ``{Cloud computing: Today and tomorrow.}'' \emph{J. Object Technol.},
  vol.~8, no.~1, pp. 65--72, 2009.

\bibitem{mao2017modnn}
J.~Mao, X.~Chen, K.~W. Nixon, C.~Krieger, and Y.~Chen, ``{Modnn: Local
  distributed mobile computing system for deep neural network},'' in
  \emph{Design, Automation \& Test in Europe Conference \& Exhibition (DATE),
  2017}.\hskip 1em plus 0.5em minus 0.4em\relax IEEE, 2017, pp. 1396--1401.

\bibitem{zhao2018deepthings}
Z.~Zhao, K.~M. Barijough, and A.~Gerstlauer, ``{Deepthings: Distributed
  adaptive deep learning inference on resource-constrained iot edge
  clusters},'' \emph{IEEE Transactions on Computer-Aided Design of Integrated
  Circuits and Systems}, vol.~37, no.~11, pp. 2348--2359, 2018.

\bibitem{huang2019deepar}
Y.~Huang, F.~Wang, F.~Wang, and J.~Liu, ``{Deepar: A hybrid device-edge-cloud
  execution framework for mobile deep learning applications},'' in \emph{IEEE
  INFOCOM 2019-IEEE Conference on Computer Communications Workshops (INFOCOM
  WKSHPS)}.\hskip 1em plus 0.5em minus 0.4em\relax IEEE, 2019, pp. 892--897.

\bibitem{alwasel2021iotsim}
K.~Alwasel, D.~N. Jha, F.~Habeeb, U.~Demirbaga, O.~Rana, T.~Baker, S.~Dustdar,
  M.~Villari, P.~James, E.~Solaiman \emph{et~al.}, ``{IoTSim-Osmosis: A
  framework for modeling and simulating IoT applications over an edge-cloud
  continuum},'' \emph{Journal of Systems Architecture}, vol. 116, p. 101956,
  2021.

\bibitem{selvaraju2017grad}
R.~R. Selvaraju, M.~Cogswell, A.~Das, R.~Vedantam, D.~Parikh, and D.~Batra,
  ``{Grad-cam: Visual explanations from deep networks via gradient-based
  localization},'' in \emph{Proceedings of the IEEE international conference on
  computer vision}, 2017, pp. 618--626.

\bibitem{adebayo2018sanity}
J.~Adebayo, J.~Gilmer, M.~Muelly, I.~Goodfellow, M.~Hardt, and B.~Kim,
  ``{Sanity checks for saliency maps},'' \emph{arXiv preprint
  arXiv:1810.03292}, 2018.

\bibitem{fummi2008systemc}
F.~Fummi, D.~Quaglia, and F.~Stefanni, ``{A SystemC-based framework for
  modeling and simulation of networked embedded systems},'' in \emph{2008 Forum
  on Specification, Verification and Design Languages}.\hskip 1em plus 0.5em
  minus 0.4em\relax IEEE, 2008, pp. 49--54.

\bibitem{bullock2004hardware}
D.~Bullock, B.~Johnson, R.~B. Wells, M.~Kyte, and Z.~Li,
  ``{Hardware-in-the-loop simulation},'' \emph{Transportation Research Part C:
  Emerging Technologies}, vol.~12, no.~1, pp. 73--89, 2004.

\bibitem{simonyan2014very}
K.~Simonyan and A.~Zisserman, ``{Very deep convolutional networks for
  large-scale image recognition},'' \emph{arXiv preprint arXiv:1409.1556},
  2014.

\bibitem{liu2021toward}
X.~Liu, W.~Yu, F.~Liang, D.~Griffith, and N.~Golmie, ``{Toward deep transfer
  learning in industrial internet of things},'' \emph{IEEE Internet of Things
  Journal}, vol.~8, no.~15, pp. 12\,163--12\,175, 2021.

\bibitem{ding2021repvgg}
X.~Ding, X.~Zhang, N.~Ma, J.~Han, G.~Ding, and J.~Sun, ``{Repvgg: Making
  vgg-style convnets great again},'' in \emph{Proceedings of the IEEE/CVF
  Conference on Computer Vision and Pattern Recognition}, 2021, pp.
  13\,733--13\,742.

\bibitem{he2016deep}
K.~He, X.~Zhang, S.~Ren, and J.~Sun, ``{Deep residual learning for image
  recognition},'' in \emph{Proceedings of the IEEE conference on computer
  vision and pattern recognition}, 2016, pp. 770--778.

\bibitem{tan2019efficientnet}
M.~Tan and Q.~Le, ``{Efficientnet: Rethinking model scaling for convolutional
  neural networks},'' in \emph{International conference on machine
  learning}.\hskip 1em plus 0.5em minus 0.4em\relax PMLR, 2019, pp. 6105--6114.

\bibitem{krizhevsky2009learning}
A.~Krizhevsky, G.~Hinton \emph{et~al.}, ``{Learning multiple layers of features
  from tiny images},'' 2009.

\bibitem{kingma2014adam}
D.~P. Kingma and J.~Ba, ``{Adam: A method for stochastic optimization},''
  \emph{arXiv preprint arXiv:1412.6980}, 2014.

\bibitem{deng2009imagenet}
J.~Deng, W.~Dong, R.~Socher, L.-J. Li, K.~Li, and L.~Fei-Fei, ``{Imagenet: A
  large-scale hierarchical image database},'' in \emph{2009 IEEE conference on
  computer vision and pattern recognition}.\hskip 1em plus 0.5em minus
  0.4em\relax Ieee, 2009, pp. 248--255.

\end{thebibliography}

\end{document}